\documentclass[aip,jcp,amsmath,amssymb,reprint]{revtex4-1}

\usepackage{hyperref}
\pdfoutput=1
\usepackage{graphicx}
\usepackage{amssymb}
\usepackage{amsmath}
\usepackage{blkarray}
\usepackage{multirow}
\usepackage{natbib}
\usepackage{epstopdf}
\usepackage{float}

\usepackage{graphicx}
\usepackage{dcolumn}
\usepackage{bm}

\usepackage{tabu} 

\begin{document}

\title{Time-independent free energies from metadynamics via Mean Force Integration}
\author{Veselina Marinova}%
\affiliation{Thomas Young Centre and Department of Chemical Engineering, University College London, London WC1E 7JE, UK.}%
\author{Matteo Salvalaglio}%
\email{m.salvalaglio@ucl.ac.uk}
\affiliation{Thomas Young Centre and Department of Chemical Engineering, University College London, London WC1E 7JE, UK.}%
\date{\today}

\begin{abstract}
Inspired by thermodynamic integration, we propose a method for the calculation of time-independent free energy profiles from history-dependent biased simulations via Mean Force Integration (MFI). MFI circumvents the need for computing the ensemble average of the bias acting on the system \textit{c(t)} and can be applied to different variants of metadynamics. Moreover, MFI naturally extends to aggregate information obtained from independent metadynamics simulations, allowing to converge free energy surfaces without the need to sample recrossing events in a single continuous trajectory. We validate MFI against one and two-dimensional analytical potentials and by computing the conformational free energy landscape of ibuprofen in the bulk of its most common crystal phase.
\end{abstract}

                          
\maketitle

\section{Introduction\label{sec:intro}}

Several enhanced sampling molecular simulation methods are aimed at computing free energy surfaces (FES) as a function of low-dimensional sets of collective variables (CVs). A strategy common to several of these methods is the introduction of an artificial bias potential that perturbs the Hamiltonian of the system, enhances the sampling of rare transitions, and facilitates the exploration of high energy regions of phase space.\cite{chipot2007free,torrie1977nonphysical,marsili2006self,grubmuller1995predicting,voter1997hyperdynamics,darve2001calculating,Laio2002,Barducci2008}. 
Metadynamics \cite{Barducci2011} is a method that implements this concept by introducing a history-dependent bias potential, iteratively updated as a sum of Gaussian contributions defined in the space of CVs. As discussed in the original publications \cite{Barducci2008,Laio2002} and in several reviews on the topic\cite{Barducci2011,valsson2016enhancing,laio2008metadynamics}, the FES recovered from the metadynamics bias potential cannot be considered inherently time-independent. 
The time-dependence of a FES computed from metadynamics is captured by the work performed by the metadynamics algorithm\cite{Tiwary2015A}, usually indicated as the time-dependent constant  $c(t)=\langle{V(\mathbf{s},t)}\rangle$\cite{Barducci2008,bonomi2009reconstructing,Tiwary2015A}. A popular approach at the calculation of time-independent free energy landscapes from time-dependent metadynamics simulations has been introduced by Tiwary and Parrinello \cite{Tiwary2015A}, who proposed an explicit expression for $c(t)$, particularly suited to the analysis of well-tempered metadynamics\cite{Barducci2008} calculations. 

Here we propose a different approach to tackle this problem, based on the observation that while promoting the exploration of phase space, metadynamics probes the gradient of the free energy hypersurface projected in CV space. 
By analysing metadynamics simulations from this perspective we propose a method for the calculation of free energy surfaces through mean force integration (MFI). 
MFI applies to different variants of metadynamics, and provides a framework to consistently \emph{patch} the sampling obtained from independent simulations in a single FES, enabling an efficient use of independent, asynchronous replicas. In this paper we discuss the MFI method, we validate it against model potentials and by computing the conformational free energy landscape of ibuprofen in the bulk of its crystal phase.

\section{Theory} 
The free energy profile along a suitably defined set of collective variables $\mathbf{s(R)}$, function of the atomic coordinates $\mathbf{R}$, can be expressed as: 
\begin{equation}
F(\mathbf{s})=-\beta^{-1}\ln{p({\mathbf{s}})}
\end{equation} 
where $\beta=(k_BT)^{-1}$, in which $k_B$ is the Boltzmann constant and $T$ is the temperature; and $p({\mathbf{s}})$ is the equilibrium probability density projected on $\mathbf{s}$. Under the effect of a perturbation of the system's Hamiltonian introduced by the bias potential $V(\mathbf{s})$, the unperturbed free energy profile $F(\mathbf{s})$ is \cite{Zwanzig1954}: 
\begin{equation}
F(\mathbf{s})=-\beta^{-1}\ln{p^b(\mathbf{s})}-V(\mathbf{s})-\langle{V(\mathbf{s})}\rangle_u
\label{eq:free_energy}
\end{equation}
Where $p^b(\mathbf{s})$ represents the equilibrium probability density under the effect of the bias potential $V(\mathbf{s})$, and $\langle{V(\mathbf{s})}\rangle_u$ is the ensemble average of the bias in the unperturbed ensemble: 
\begin{equation}
\langle{V(\mathbf{s})}\rangle_u=\beta^{-1}\ln\frac{\int_\Omega{e^{-\beta{F(\mathbf{s})+\beta{V(\mathbf{s})}}}}d\mathbf{s}}{{\int_{\Omega}{e^{-\beta{F(\mathbf{s})}}d\mathbf{s}}}}
 \label{eq:constant}
\end{equation}

It should be noted that Eq. \ref{eq:free_energy} provides an implicit expression for $F(\mathbf{s})$, which appears on the right hand side within $\langle{V(\mathbf{s})}\rangle_u$. In Eq. \ref{eq:free_energy} the term $\langle{V(\mathbf{s})}\rangle_u$ is \emph{non-local}, i.e. it contributes to the absolute value of $F(\mathbf{s})$, but the ensemble average operation makes it independent with respect to $\mathbf{s}$. The calculation of this term is essential in Umbrella Sampling (US) \cite{haydock1990tryptophan}, for estimating free energy profiles from multiple biased simulations that sample different regions of $\mathbf{s}$. The estimate of this term in US simulations is commonly carried out iteratively via the Weighted Histogram Analysis Method (WHAM) algorithm\cite{roux1995calculation,kumar1992weighted}. 
In the context of adaptive enhanced sampling methods such as metadynamics, in which the bias potential changes in time according to the sampling history of the system, i.e. $V(\mathbf{s},t)$, the term $\langle{V(\mathbf{s})}\rangle_u$ is a function of time and corresponds to the constant $c(t)$ for which an explicit formulation has been proposed by Tiwary and Parrinello\citep{Tiwary2015A}.
In the following we illustrate how MFI does not require an estimate of $c(t)$ to obtain a time-independent estimate of $F(\mathbf{s})$.

We shall begin by noting that in metadynamics the bias potential $V(\mathbf{s},t)$ is evolved \emph{discretely} in time, through updates performed at regular time intervals of length $\tau$. 

Between two consecutive updates of the bias potential, performed at times $t$ and $t+\tau$, the system evolves under the effect of the stationary bias $V_t(\mathbf{s})$ and samples the biased probability density $p_t^b(\mathbf{s})$.
The estimate of $p_t^b(\mathbf{s})$ obtained during the sampling time $\tau$ is typically localised in a small subregion of $\mathbf{s}$, moreover in different iterations of the bias update algorithm, the perturbation of the Hamiltonian introduced by the biasing potential $V_t(\mathbf{s})$ is different. Hence, in order to reconstruct a global free energy surface $F(\mathbf{s})$ from Eq. \ref{eq:free_energy}, the term $\langle{V_t(\mathbf{s})}\rangle_u$ is necessary and has to be evaluated at every update of the bias potential. 

In MFI we approach this problem by taking inspiration from the Umbrella Integration (UI) method\cite{Kastner2005}. In UI, instead of straightforwardly applying Eq. \ref{eq:free_energy}, the estimate of the non-local term $\langle{V(\mathbf{s})}\rangle_u$ is circumvented by computing the \emph{mean force} in CV space, $\nabla{F(\mathbf{s})}$. The free energy surface $F(\mathbf{s})$ is then obtained by numerical integration of the mean force. 

In order to gradually introduce complexity, in the following section we shall outline the details of the method for a mono-dimensional CV space. We then discuss the generalisation of MFI to CV spaces of higher dimensionality, and finally we outline how MFI provides the means to consistently merge the sampling obtained from independent simulations into a single estimate of the free energy surface.\\

\paragraph*{Mean Force Integration in 1D CV spaces}
For the sake of clarity, let us begin by considering a simple case in which $\mathbf{s}$ is a mono-dimensional CV space, thus indicated as the scalar $s$. The derivative of the free energy profile with respect to $s$ is: 
\begin{equation}
\frac{dF_t(s)}{d{s}}=-\frac{d\beta^{-1}\ln{p_t^b(s)}}{d{s}}-\frac{d{V_t(s)}}{d{s}}
\label{eq:working_eq}
\end{equation}

where $\frac{dF_t(s)}{d{s}}$ is the mean force in CV space obtained from the sampling performed in the time interval $[t;t+\tau]$, the term $\frac{d{V_t(s)}}{d{s}}$ is the derivative of the bias potential updated at time $t$, which is stationary during the time interval $[t;t+\tau]$. Finally, the term $\frac{d\beta^{-1}\ln{p_t^b(s)}}{d{s}}$ corresponds to the mean force in $s$ under the effect of the perturbation due to the bias potential $V_t(s)$, sampled during the time interval $[t;t+\tau]$. 
It should be noted that the term $\frac{d{V_t(s)}}{d{s}}$ is accumulated from all the updates of the bias potential performed up to time $t$. On the contrary, the term associated with $p_t^b(s)$ is estimated anew after every iterative update of the bias potential.

During a metadynamics simulation the bias is updated frequently, usually in thousands of iterations. Each update of the bias potential will yield a mean force estimate $\frac{dF_t(s)}{d{s}}$. 
Following the approach proposed in Umbrella Integration\citep{Kastner2005}, the average mean force realization at time $t$ is estimated as: 
\begin{equation}
\bigg\langle{\frac{dF_t(s)}{d{s}}}\bigg\rangle_t=\frac{\sum_{t^{\prime}=1}^{t} p_{t^\prime}^b(s) \frac{{dF_{t^\prime}(s)}}{d{s}}}{\sum_{t^{\prime}=1}^{t} p_{t^\prime}^b(s)}
\label{eq:average_force}
\end{equation}

From $\big\langle{\frac{dF_t(s)}{d{s}}}\big\rangle_t$ a time-independent estimate of $F(s)$ is obtained through numerical integration. It should be noted also that we indicate with $\langle{...}\rangle_t$ the estimate at time $t$ of the mean force in $s$, however $\langle{\frac{dF_t(s)}{d{s}}}\rangle_t$ is an inherently time-independent quantity.

In order to apply Eq. \ref{eq:average_force}, we derive an analytical expression for the terms $\frac{d{V_t(s)}}{d{s}}$ and $\frac{d\beta^{-1}\ln{p_t^b(s)}}{d{s}}$. 

The former can be straightforwardly computed as the derivative of the sum of Gaussians accumulated up to time $t$: 
\begin{equation}
\frac{d{V_t(s)}}{d{s}}={\sum_{t^{\prime}=1}^{t}{-\frac{w_t(s-s_{t^{\prime}})}{{\sigma_{M,t}}^2}}\exp\left[{-\frac{1}{2}\frac{\left({s-s_{t^{\prime}}}\right)^2}{{\sigma_{M,t}}^2}}\right]}
\label{eq:force1}
\end{equation}
where $w_t$, and $\sigma_{M,t}$ are the values of the Gaussian height and width at time $t$, and $s_t$ is the position in CV space that corresponds to the mean value of the Gaussian deposited at time $t$.  

As demonstrated in the results section, this expression holds regardless of the protocol followed to update $V_t(\mathbf{s})$, and is therefore applicable to any metadynamics variant including standard MetaD ($w_t$, and $\sigma_{M,t}$ constant), WTmetaD and TTmetaD \cite{Dama2014}($w_t$, updated at every iteration, $\sigma_{M,t}$ constant), and adaptive Gaussians metaD\cite{Branduardi2012}($\sigma_{M,t}$ updated at every iteration). 

In order to express the term $\frac{d\beta^{-1}\ln{p_t^b(\mathbf{s})}}{d{s}}$ in a general form, we apply a kernel density estimation of $p_t^b(\mathbf{s})$, the biased probability density sampled in the time interval $[t;t+\tau]$. Using Gaussian kernels $p_t^b(\mathbf{s})$ takes the form: 
\begin{equation}
p_t^b(s)={\frac{1}{n_{\tau}h\sqrt{2\pi}}}\sum_{t^\prime=t}^{t+\tau} \exp\left[{-\frac{\left({s-s_{t^\prime}}\right)^2}{2h^2}}\right]
\label{eq:biasprob}
\end{equation}
where $n_{\tau}$ is the number of frames sampled in the time interval $[t;t+\tau]$, $h$ is the kernel bandwidth, $s_t$ is the instantaneous value of $\mathbf{s}$. Thus, the mean force contribution associated with the biased probability density term is: 
\begin{equation}
\frac{d\beta^{-1}\ln{p_t^b(s)}}{d{s}}=\frac{{\sum_{t^\prime=t}^{t+\tau}-\frac{s-s_{t^\prime}}{\beta{h^2}}}\exp\left[{-\frac{\left({s-s_{t^\prime}}\right)^2}{2h^2}}\right]}{\sum_{t^\prime=t}^{t+\tau} \exp\left[{-\frac{\left({s-s_{t^\prime}}\right)^2}{2h^2}}\right]}
\label{eq:force2}
\end{equation}

We note that in order to apply MFI to metadynamics we need to significantly depart from the hypothesis of Kastner et al. in Umbrella Integration, i.e. of $p_t^b(\mathbf{s})$ being a mono-modal probability density, normally distributed around the average value of $\mathbf{s}$. This hypothesis holds for the Umbrella Sampling protocol, where the bias potential confines sampling in a specific region of $\mathbf{s}$, localised around a certain target position. However it breaks down in the case of metadynamics, where each repulsive Gaussian contribution tends to push the system away from its center. 
This yields biased distributions that are far from being mono-modal, even on the short timescale of $\tau$, which can nevertheless be faithfully captured by Eq. \ref{eq:biasprob}.

\begin{figure*}[ht]
\includegraphics[width=0.75\linewidth]{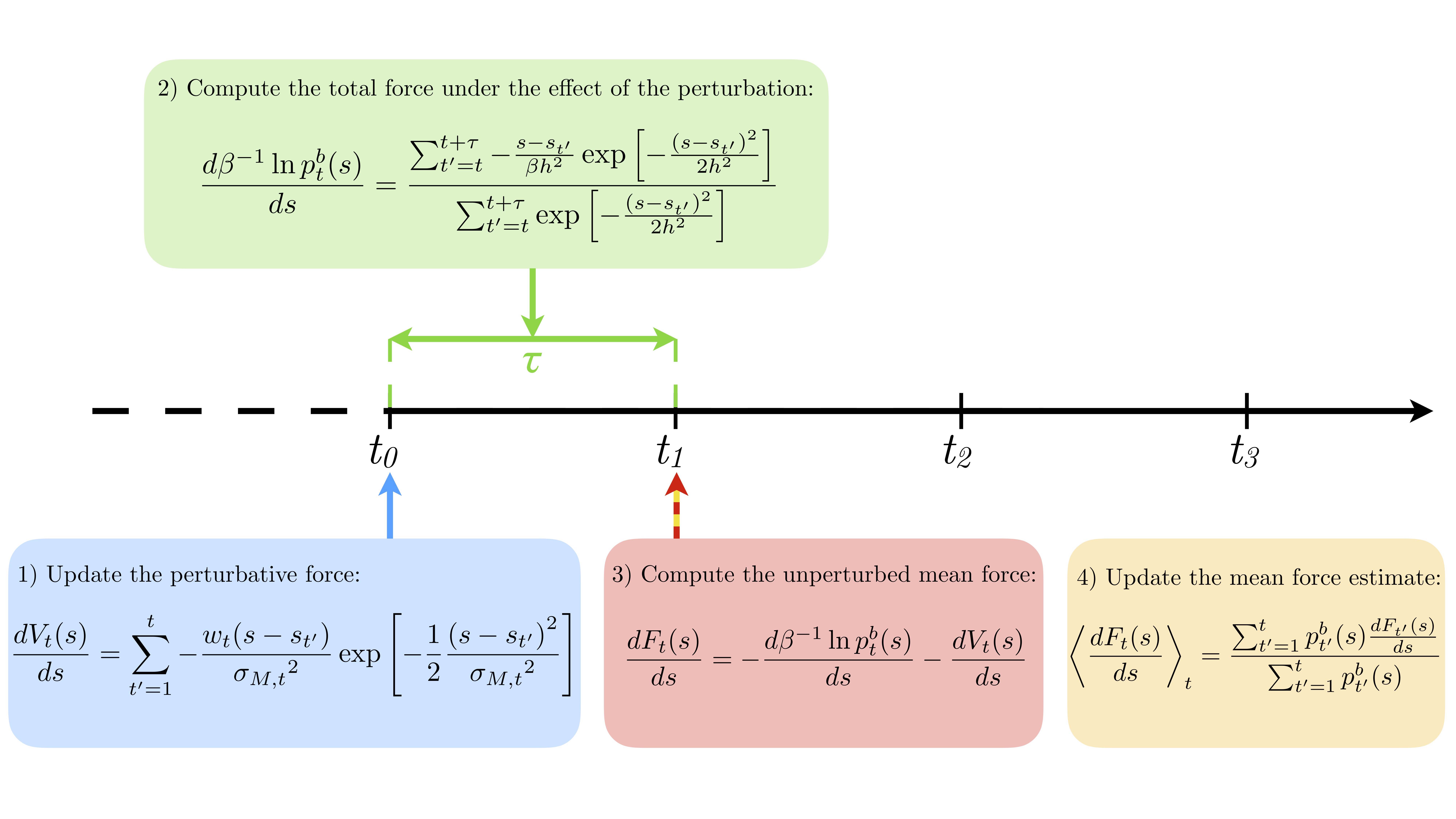}
\caption{A graphical representation of the MFI algorithm. At instants of time $t_0, t_1,...,t_n$ the bias is updated with the addition of a new Gaussian. The perturbative force $\frac{dV}{ds}$ is therefore consistently updated at every time step $t_0, t_1,...,t_n$. In the interval of time $\tau$, the total thermodynamic force, active under perturbation is estimated from direct sampling of the biased probability density, obtained under the effect of bias $V_t(s)$. At the end of the interval $\tau$, the unperturbed mean force for the time interval $t_i+\tau$ is estimated, and the total mean force estimate is updated.} 
\label{fig:algorithm}
\end{figure*}

Combining Eq. \ref{eq:working_eq}-\ref{eq:force2} we obtain an analytic expression for the mean force in $\mathbf{s}$: 
\begin{widetext}
\begin{equation}
\bigg\langle{\frac{dF_t(s)}{d{s}}}\bigg\rangle_t=\frac{1}{{\sum_{t^{\prime}=1}^{t} p_{t^\prime}^b(s)}}\left\{  {\sum_{t^{\prime}=1}^{t}\sum_{t^{\prime\prime}=t^\prime}^{t^\prime+\tau}\frac{s-s_{t^{\prime\prime}}}{\beta\,n_\tau{h^3}\sqrt{2\pi}}\exp\left[{-\frac{\left({s-s_{t^{\prime\prime}}}\right)^2}{2h^2}}\right]}+{\sum_{t^{\prime}=1}^{t} p_{t^\prime}^b(s) \frac{{dV_{t^\prime}(s)}}{d{s}}}\right\}
\label{eq:final_force_1D}
\end{equation}
\end{widetext}
A graphical scheme representing the calculation procedure for the update of the mean force through Eq. \ref{eq:final_force_1D} is reported in Fig. \ref{fig:algorithm}.
It should be noted that, while the second term of Eq. $\ref{eq:final_force_1D}$ depends on the specific bias protocol, the first term is generally valid for any history-dependent biasing protocol based on discrete iterative updates of the bias potential. By numerically integrating $\big\langle{\frac{dF_t(\mathbf{s})}{d{\mathbf{s}}}}\big\rangle_t$ one can obtain a time-independent estimate of the free energy surface $F(\mathbf{s})$. 
In the long time limit $p_t^b(\mathbf{s})$ approaches the limit distribution associated with the chosen sampling method. In the case of standard metadynamics, in the long time limit the biased distribution becomes flat, and the term $\frac{d\beta^{-1}\ln{p_t^b(\mathbf{s})}}{d{\mathbf{s}}}\rightarrow{0}$, thus recovering the standard estimator of the free energy $F(\mathbf{s})=-V(\mathbf{s})+C$.
If needed, by using the integrated profile $F(\mathbf{s})$ in Eq. \ref{eq:constant} one can compute the non-local time dependent constant $c(t)=\langle{V(\mathbf{s},t)}\rangle_u$ and perform on-the-fly reweighting for additional variables function of the system's coordinates $O(\mathbf{R})$ as: 
\begin{equation}
\langle{(\mathbf{R})}\rangle_u =\langle{O(\mathbf{R})\exp\left[\beta{V_t}(\mathbf{s})-\beta\langle{V_t(\mathbf{s})}\rangle_u\right]}\rangle_t
\label{eq:reweight}
\end{equation}
where, following the notation of Ref.\cite{Tiwary2015A}, with angular brackets on the right-hand side we indicate an average over the biased simulation. In contrast to the approach of Tiwary and Parrinello we do not invoke any assumption on the bias evolution, and Eq. \ref{eq:reweight} is valid for any biasing protocol, as long as the bias is updated at discrete time intervals of length $\tau$ that enable the local estimate of $p_t^b(\mathbf{s})$ under the effect of a stationary bias $V_t(\mathbf{s})$. \\

\paragraph*{Generalization to a $d$-dimensional free energy surface}
The result obtained for the monodimensional case can be straightforwardly generalised to the calculation of an arbitrary-dimensional free energy hypersurface $\mathbf{s}$, where  Eq. \ref{eq:working_eq} becomes \cite{Kastner2009}: 
\begin{equation}
\nabla{F_t(\mathbf{s})}=-\beta^{-1}\nabla{\ln{p_t^b(\mathbf{s})}}-\nabla{V_t(\mathbf{s})}
\label{eq:working_eq_ND}
\end{equation}

In this case, the $d$-dimensional biased probability distribution $p_t^b(\mathbf{s})$ sampled in the time interval $\tau$ can be obtained with a multivariate kernel density estimation. Using multivariate Gaussian kernels $p_t^b(\mathbf{s})$ takes the form: 
\begin{multline}
p_t^b(\mathbf{s})=\frac{1}{n_\tau\left(2\pi\right)^{d/2}\vert{\mathbf{h}}\vert^{1/d}}  \times \\ \times \sum_{t^\prime=t}^{t+\tau}\exp\left[-\frac{1}{2}\left(\mathbf{s}-\mathbf{s}_{t^\prime}\right)^\intercal\mathbf{h}^{-2}\left(\mathbf{s}-\mathbf{s}_{t^\prime}\right)\right]
\end{multline}

where $d$ is the dimensionality of the CV space $\mathbf{s}$, $\mathbf{h}$ is the variance/covariance matrix of the multivariate Gaussian kernel, and $\vert{\mathbf{h}}\vert$ is the determinant of the variance/covariance matrix. For the typical case of $d$=2, and diagonal $\mathbf{h}$ we have that $\vert{\mathbf{h}}\vert^{1/d}=\sqrt{h_1h_2}$, and $\left(\mathbf{s}-\mathbf{s}_{t^\prime}\right)^\intercal\mathbf{h}^{-2}\left(\mathbf{s}-\mathbf{s}_{t^\prime}\right)=\sum_{i=1}^2\left(\frac{s_i-s_{i,{t^\prime}}}{h_i}\right)^2$ where $h_1$ and $h_2$ are the bivariate Gaussian kernel bandwidths in dimensions 1 and 2 respectively. 

 In a $d$-dimensional metadynamics simulation the bias $V_t(\mathbf{s})$ is represented as the sum of multivariate Gaussian contributions in $d$ dimensions. It should be noted that while in standard and well-tempered metadynamics the variance/covariance matrix is kept constant, in the case of adaptive Gaussians metaD, the variance/covariance matrix of the multidimensional Gaussian kernels is not diagonal and its terms are adaptively estimated based on sampling. 

\begin{figure*}[ht]
\includegraphics[width=1\linewidth]{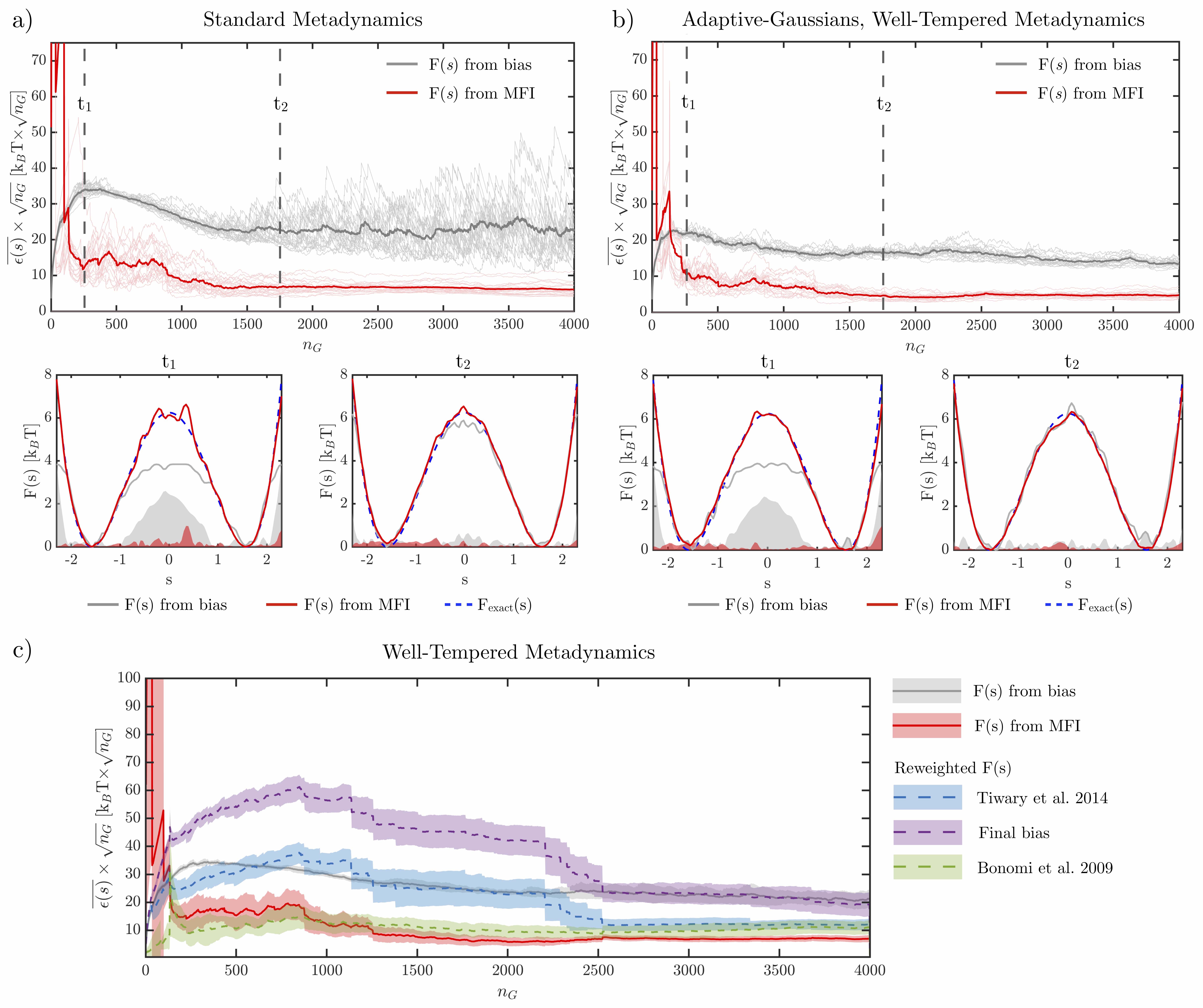}
\caption{Convergence analysis of MFI for a double well potential of the form: $F_{exact}(\mathbf{s})=-5s^2+s^4$. a-c) Plot of $\overline{\epsilon{(s)}}\times{\sqrt{n_G}}$ as a function of the number of Gaussian updates of the bias potential $n_G$. The mean of thirty independent trajectory is highlighted with a thicker solid line. a) Standard metadynamics. b) Adaptive Bias metadynamics. c) In the case of Well Tempered metadynamics we report a comparison between MFI and three typical reweighting methods\cite{Gimondi2018}. For the sake of clarity in this panel we report the average of 30 independent simulations as a solid line, and the standard deviation as shaded area.  As it can be seen, for the simple double well potential studied in this section, MFI displays the faster asymptotic convergence rate.}
\label{fig:1Dpotential}
\end{figure*}

\section{Results and Discussion}
In this section we assess the accuracy of MFI by computing the free energy surface in the case of one and two dimensional model potentials using metadynamics. After that we demonstrate, for the case of ibuprofen conformational isomerism in its crystal bulk, how MFI enables the efficient and accurate calculation of free energy surfaces by merging the information obtained from independent simulations that do not include recrossing events. 

All metadynamics simulations analysed in this section, both on analytical model potentials and on ibuprofen, were performed using PLUMED 2.4 \cite{tribello2014plumed}.  
In the case of ibuprofen GROMACS 5.1.4 \cite{VanDerSpoel2005} was used as a molecular dynamics engine with the Generalised Amber Force Field (GAFF) used to define the potential energy of ibuprofen \cite{Wang2004,Marinova2018}. A detailed description of the simulations setup for ibuprofen conformational isomerism in the crystal bulk can be found in Supplementary Materials, as well as in Ref. \cite{Marinova2018}.  All the data and PLUMED input files required to reproduce the results reported in this paper are available on PLUMED-NEST (www.plumed-nest.org), the public repository of the PLUMED consortium \cite{NEST}, as plumID:19.071.

\subsection*{Model Potentials}
\paragraph*{1D double well.} In order to quantify the accuracy and the convergence rate of the free energy surface obtained by MFI we start from a simple 1D double well model potential.  We perform Langevin dynamics, biased with different metadynamics variants, i.e. standard, well-tempered, transition-tempered and adaptive Gaussians.

In Fig.\ref{fig:1Dpotential}(a-c) we report, as a function of the number of Gaussians added to define the bias potential, $n_G$, the quantity $\overline{\epsilon{(s)}}\times{\sqrt{n_G}}$. Where  $\overline{\epsilon{(s)}}$ is the average absolute error in the estimate of the free energy profile $F(s)$. The exact free energy profile is defined by the expression $F_{exact}(\mathbf{s})=-5s^2+s^4$. 
In the absence of systematic errors $\overline{\epsilon{(s)}}\times{\sqrt{n_G}}$ converges to a flat plateau\cite{Barducci2008}. As can be seen in panels $a$-$c$ such condition, is met in plain metadynamics ($a$), Well Tempered Adaptive Bias metadynamics ($b$), and Well Tempered metadynamics ($c$). The relative position of the horizontal plateau for the quantity $\overline{\epsilon{(s)}}\times{\sqrt{n_G}}$ allows to compare the relative rate of convergence of different methods. For instance, the lower the plateau, the faster the convergence. From panels $a$ and $b$ one can see that MFI (red) provides faster convergence compared to the standard estimator based on the total bias deposited (gray) for both standard metadynamics and adaptive bias metadynamics. Similar results are reported in the Supplementary material for Well Tempered and Transition Tempered Metadynamics.

Focussing on Fig.\ref{fig:1Dpotential}(a,b), and in particular on the free energy profiles obtained at $t_1$ and $t_2$ (indicated with dashed lines in Fig. \ref{fig:1Dpotential}a) and b), one can see that the estimate of the double-well FES obtained from MFI (solid red curve) provides a more accurate estimate of the analytical FES (dashed blue line) than that obtained from the bias potential (solid gray curve) after the same number of bias updates. In shaded red (MFI) or gray (bias) is represented the position-dependent absolute error, demonstrating  that the error associated with MFI is significantly smaller than the associated with FES estimates based only on the Bias potential. 

In Fig. \ref{fig:1Dpotential}c we report for Well Tempered metadynamics, a comparison between the convergence rate of MFI and three typical reweighting methods. We consider the reweighting approaches proposed by Tiwary and Parrinello\cite{Tiwary2015A}, Bonomi et al. \cite{bonomi2009reconstructing}, and a simplistic reweighting strategy based on considering all the sampling during the WTmetaD as if it was performed under the effect of the final bias.\cite{Gimondi2018}
It can be seen that, like MFI, both the reweighting methods of Bonomi et al. \cite{bonomi2009reconstructing}, and Tiwary et al. \cite{Tiwary2015A} converge faster than estimates of the FES based solely on the bias potential. Among all methods compared in Fig. \ref{fig:1Dpotential}c, MFI shows the fastest convergence rate. 

The analysis of the results obtained for this initial test case shows that MFI provides an accurate estimates of the model FES, which rapidly converges to the analytical results once a  crossing event is sampled (i.e. around $n_G$=200 in Fig. \ref{fig:1Dpotential}a-c). 
We note finally that the rate of convergence and the quality of the FES obtained with MFI are very moderately affected by the choice of particular metadynamics variant, and that in the case of WT metadynamics, where several reweighting strategies can be compared\cite{Tiwary2015A,bonomi2009reconstructing,Gimondi2018}, MFI display the fastest asymptotic convergence rate. 

\begin{figure*}[ht]
\includegraphics[width=0.95\linewidth]{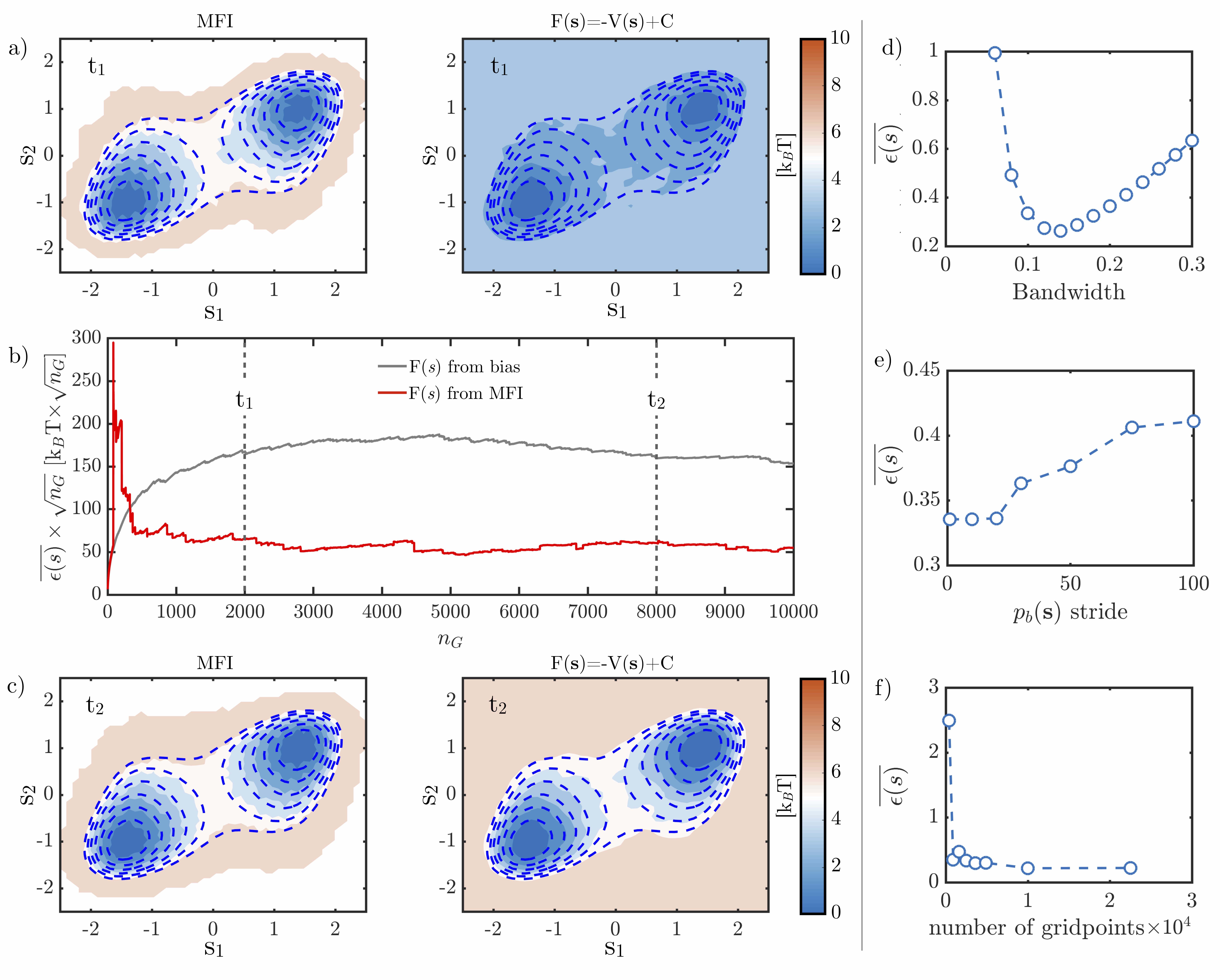}
\caption{(a-c) Convergence analysis of MFI against the standard estimators of the free energy for a standard metadynamics simulation for a 2D, double well model potential ($F_{exact}(\mathbf{s})=-3s_1^2+s_1^4-3s_1s_2+s_2^4$). In panel b the quantity $\overline{\epsilon{(\mathbf{s})}}\times{\sqrt{n_G}}$ is reported as a function of the number of bias potential updates $n_G$ for the low energy region $F(\mathbf{s})_{exact}<10$ $k_BT$. In the top and bottom panels (a and c) we compare the estimates of $F(\mathbf{s})$ computed with MFI and as $F(\mathbf{s})=-V(\mathbf{s})$ at two different times ($t_1$ and $t_2$). It can be seen that MFI provides a faster convergence of the FES estimate. (d-f) Analysis of the error associated to free parameters in Eq. \ref{eq:final_force_1D}. (d) Error dependence on the kernel bandwidth $h$ used to define $p_b(\mathbf{s})$ (f) Error dependence on the stride used to extract data-points used to define $p_b(\mathbf{s})$ (g) Error dependence on the number of gridpoints used to numerically integrate the mean force estimated through Eq. \ref{eq:final_force_1D}.}
\label{fig:2Dpotential}
\end{figure*}

\paragraph*{2D double well.} In order to further demonstrate the applicability of MFI to 2D surfaces and assess its sensitivity to key parameters appearing in the definition of the thermodynamic force expression reported in Eq. \ref{eq:working_eq_ND}, we perform Langevin dynamics simulations on a 2D double-well model potential ($F_{exact}(\mathbf{s})=-3s_1^2+s_1^4-3s_1s_2+s_2^4$). 

In Fig. \ref{fig:2Dpotential}a) and c) we report the FES obtained with MFI and compare it to the FES obtained as $F(\mathbf{s})=-V(\mathbf{s})$ at the same simulation time. It can be seen that, the FES obtained through MFI provides a better representation of the exact analytical potential given the same sampling. This is particularly evident in the transition region between the two local minima. In Fig. \ref{fig:2Dpotential}b) we show the time evolution of the quantity $\overline{\epsilon{(s)}}\times{\sqrt{n_G}}$ demonstrating that MFI does not display any systematic error, and that it converges faster than the negative of the bias potential to $F_{exact}$. The reason for the faster convergence is that, similarly to what happens to the method proposed by Tiwary et al., MFI provides an expression for the mean force that holds for any time, and not only in the long time limit. Hence realisations at short times, when the bias potential deposited in the CV space between minima is scarce, are already representative of the exact FES. 

In Fig.\ref{fig:2Dpotential}d-f), we conduct a systematic investigation of the error associated to the parameters that can be freely selected to inform Eq. \ref{eq:final_force_1D}. Such parameters are: \emph{i}) the bandwidth of the Gaussian kernels used to construct the biased probability density $p_b(\mathbf{s})$ (Fig.\ref{fig:2Dpotential}d), \emph{ii}) the stride used to extract data points from the system's evolution to build a kernel density estimator of $p_b(\mathbf{s})$ (Fig.\ref{fig:2Dpotential}e), and \emph{iii}) the number of grid points used to numerically integrate the mean force in CV space, and obtain an estimate of the FES (Fig.\ref{fig:2Dpotential}f). 

In all cases the dependence of the error on parameters is weak, with a mean absolute error of the order of $k_BT$ over the entire of the parameter space investigated. Nevertheless, the dependence of the error on each of the three parameters is different. For instance, the error dependence on the number of grid-points and on M the stride display trends typical of numerical convergence, in which the error decreases monotonically with a smaller stride (i.e. more data points) and a finer grid used for numerical integration. The error dependence on the bandwidth used to estimate $p_b(\mathbf{s})$ instead displays a non-monotonic behaviour, which for the 2D Langevin simulation has a minimum for a bandwidth comprised between 0.1 and 0.2. This range corresponds to half of the mean fluctuation of the CV in the time interval $\tau$, that separates successive updates of the metadynamics potential.  This observation confirms the validity of the heuristics typically implemented in the selection of the Gaussian width in setting up a metadynamics simulation also in defining a sensible bandwidth for the calculation of the FES through MFI. 

\begin{figure*}[ht]
\includegraphics[width=1.0\linewidth]{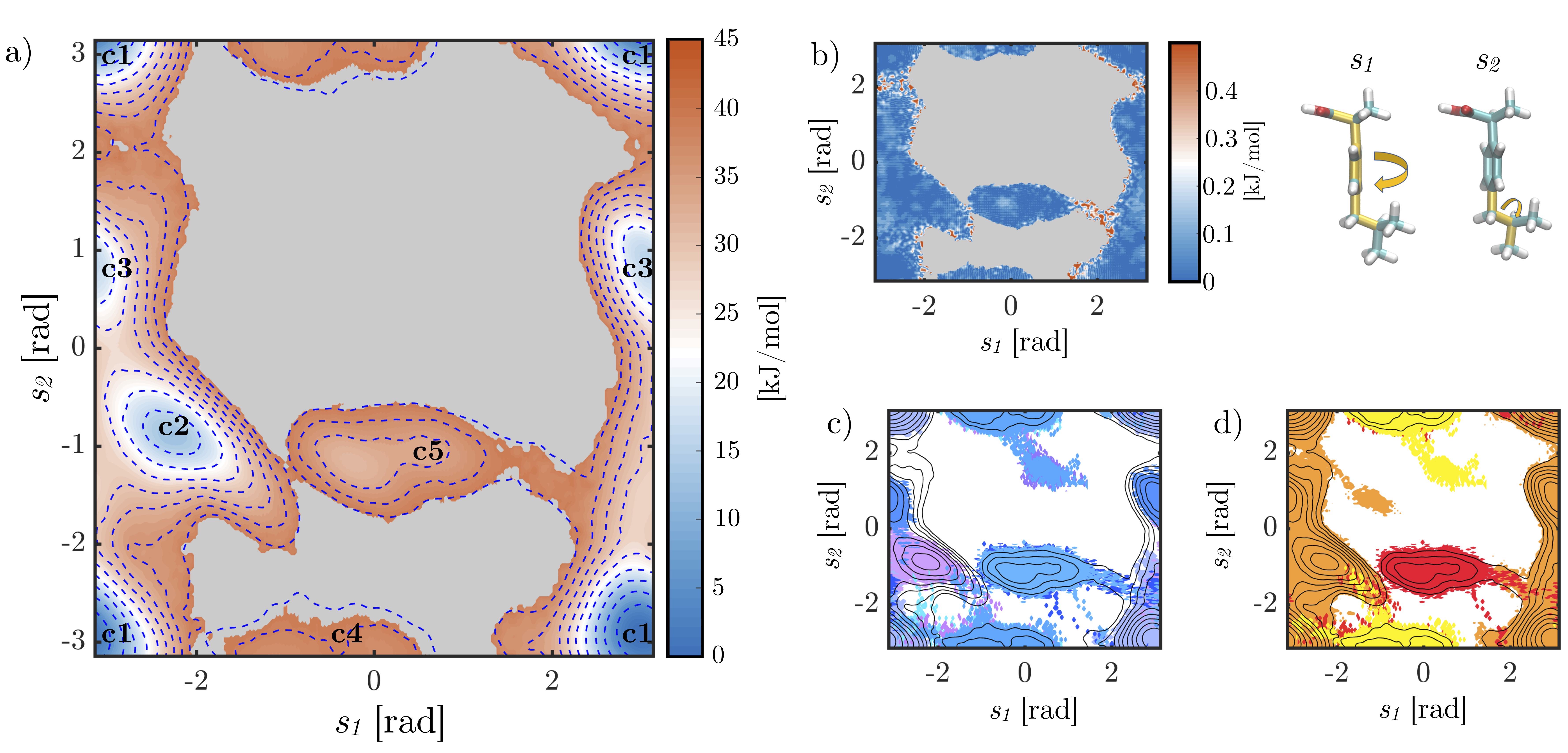}
\caption{a) FES (blue to red colourmap) of ibuprofen conformational rearrangement in the crystal bulk recovered with MFI from independent simulations. Isocontours (blue dashed line) represent the free energy profile obtained from a 120 ns WTmetaD simulation reported in SI. b) The position-dependent error of the FES reported in a). c) CV space explored using simulation sets 1 to 6 as shown in Tab. \ref{tab:SimulationSets} d) CV space explored using simulation set 1 to 10 as reported in Tab. \ref{tab:SimulationSets} }
\label{fig:ibuprofenpatching}
\end{figure*}

\paragraph*{Patching independent metadynamics simulations: the case of ibuprofen.} 
To demonstrate the effectiveness of MFI in making the most of the sampling carried out by independent simulations, we compute the FES associated with ibuprofen conformational rearrangements in the crystal bulk. 
Ibuprofen is a small organic molecule consisting of a phenyl ring with two \textit{para}-substituents and a chiral centre. 
In a recent work\cite{Marinova2018}, we have demonstrated how, unlike in solution, when ibuprofen is embedded in the crystal bulk its conformational rearrangement is restricted and the escape from its crystallographic conformational state (named c1 in Fig. \ref{fig:ibuprofenpatching}a) is a rare event with an associated timescale of around 100 ns. 

In this work, we compute the free energy landscape associated to ibuprofen in the crystal bulk by performing a series of independent metadynamics simulations initialized in each of its conformational states. 
Simulations are stopped once a prescribed crossing event is observed. The sets of simulations, together with their initial and final configuration, and average duration are reported in Tab. \ref{tab:SimulationSets}. Additional details on the configurational landscape and the simulation setup are provided as Supplementary Material. 

\begin{table}[h]
\caption{Simulation settings used for the calculation of the FES of ibuprofen conformational isomerism using MFI.}
	\begin{tabular}{ c | c | c | c }
		 \textbf{Set} & \textbf{Initial state} & \textbf{Final State} & \textbf{Average Length} \\
		 &  &  & [ns] \\
		\hline
		1 & c1 & any other & 1.7  \\		
		2 & c2 & any other & 0.05 \\ 
		3 & c3 & any other & 0.013 \\
		4 & c4 & any other & 0.015 \\
		5 & c5 & any other & 0.05 \\
		6 & c6 & any other & 0.0008 \\
		7 & c1 & c4,c5 or c6 & 15 \\
	    8 & c4 & c1,c2 or c3 & 0.13\\
	    9 & c5 & c1,c2 or c3 & 0.05\\
	   10 & c6 & c1,c2 or c3 & 0.009\\
	\end{tabular}
    \label{tab:SimulationSets}
\end{table}

In order to compute the sampling error associated to MFI we divide the simulations in six groups, each containing five randomly selected simulations from every set reported in Tab. \ref{tab:SimulationSets}.
In Fig. \ref{fig:ibuprofenpatching}a) the FES (blue to red colormap) obtained by averaging the results of each group is reported. The FES is represented in the space of two torsional angles. A \textit{global} one, $s_1$, which describes the rearrangement of the \textit{para}-substituents of the phenyl ring, and a \textit{local} one, $s_2$, capturing the rotation of the methyl groups within the isobutanyl substituent\cite{Marinova2018}. Starting from the crystallographic conformer, c1, the rotation of the local torsional angle generates isomers c2 and c3.  Rotation along the global torsional angle in each of the conformers c1, c2 and c3 results in respectively conformers c4, c5 and c6 as shown in Fig. 3 in the SI.
In Fig. \ref{fig:ibuprofenpatching}a) we also report, as a term of comparison, the isocontours of a reference FES obtained with standard post-processing of a 120 ns long WTmetaD simulation performed as reported in the SI. The position-dependent standard error in the FES computed with MFI is generally rather small, as shown in Fig. \ref{fig:ibuprofenpatching}b), with a maximum of 0.5 kJ/mol and an average of $\sim0.1$ kJ/mol in the region of interest ($\Delta{F}<45$ kJ/mol). 

We note that the transition pathways between states need to be sufficiently sampled to obtain an accurate estimate of the free energy difference between them. In Fig. \ref{fig:ibuprofenpatching}c) we show the sampling achieved when using simulation sets 1 to 6, which correspond to simulations that are stopped as soon as the starting configuration transforms in any of the other stable conformers. On the plot each set is represented in a different colour, according to its starting configuration. These sets of simulations allow an accurate reconstruction of the free energy profile for the CV space occupied by conformers c1, c2 and c3 as the sampling in the channels between them is sufficient to connect the corresponding regions of the free energy, but are insufficient to generate the full FES accurately. 

By including sets 7 to 10 the sampling in the transition channels c2 $\Longleftrightarrow$ c4 and c5 $\Longleftrightarrow{}$ c1 is improved by forcing a cross over along the $s_1$ direction at shown in In Fig. \ref{fig:ibuprofenpatching}d) yielding a fully converged FES. 

By employing MFI we have successfully reconstructed an accurate FES associated with the conformational rearrangement of ibuprofen in the crystal bulk from independent simulations without recrossings. 
MFI proves to be a powerful tool in obtaining free energy profiles without the need of sampling recrossing transitions along the same, continuous trajectory. 

\section{Conclusions}
In this work, we have introduced MFI as a method for the calculation of time-independent free energy surfaces from history-dependent metadynamics simulations. Inspired by Umbrella integration, MFI is based on the analytic evaluation of the mean force in CV space and does not require the explicit calculation of the ensemble average of the deposited bias $\langle{V(\mathbf{s})}\rangle$. MFI applies to any history-dependent biasing schedule, provided that the bias is updated in discrete time steps, separated by a time interval $\tau$. We have shown that MFI provides accurate and rapidly converging estimates of analytical free energy profiles in one and two dimensions. Furthermore, we have demonstrated the applicability of MFI to the calculation of free energy surfaces from ensembles of independent metadynamics trajectories without the requirement of sampling recrossing events within the same continuous trajectory. We anticipate that MFI will be useful to obtain and systematically refine FES when metadynamics trajectories independently sampling realizations of a rare event are available. An example of this application would be the recovery of FES from ensembles of biased trajectories generated with an infrequent metadynamics protocol.\cite{salvalaglio2014assessing,tiwary2013metadynamics}

\subsection*{Supplementary Material}
Convergence for 1D model potential in the case of Well Tempered and Transition Tempered Metadynamics, additional information on the Ibuprofen simulation setup, and on the associated free energy landscape, including convergence and error analysis.  

\subsection*{Acknowledgements}
This work was financially supported by Pfizer, and by the Engineering and Physical Sciences Research Council (EPSRC) grant EP/R018820/1. We acknowledge the Legion High Performance Computing Facility for access to Legion@UCL and associated support services in the completion of this work. 

\section{References}
\bibliographystyle{unsrt}
\bibliography{refs}
\end{document}


\title{Time-independent free energies from metadynamics via Mean Force Integration - Supplementary Material}
\author{Veselina Marinova}%
\affiliation{Thomas Young Centre and Department of Chemical Engineering, University College London, London WC1E 7JE, UK.}%
\author{Matteo Salvalaglio}%
\email{m.salvalaglio@ucl.ac.uk}
\affiliation{Thomas Young Centre and Department of Chemical Engineering, University College London, London WC1E 7JE, UK.}%
\date{\today}

\maketitle

\section*{Convergence 1D potential}
\begin{figure*}[ht]
\includegraphics[width=0.98\linewidth]{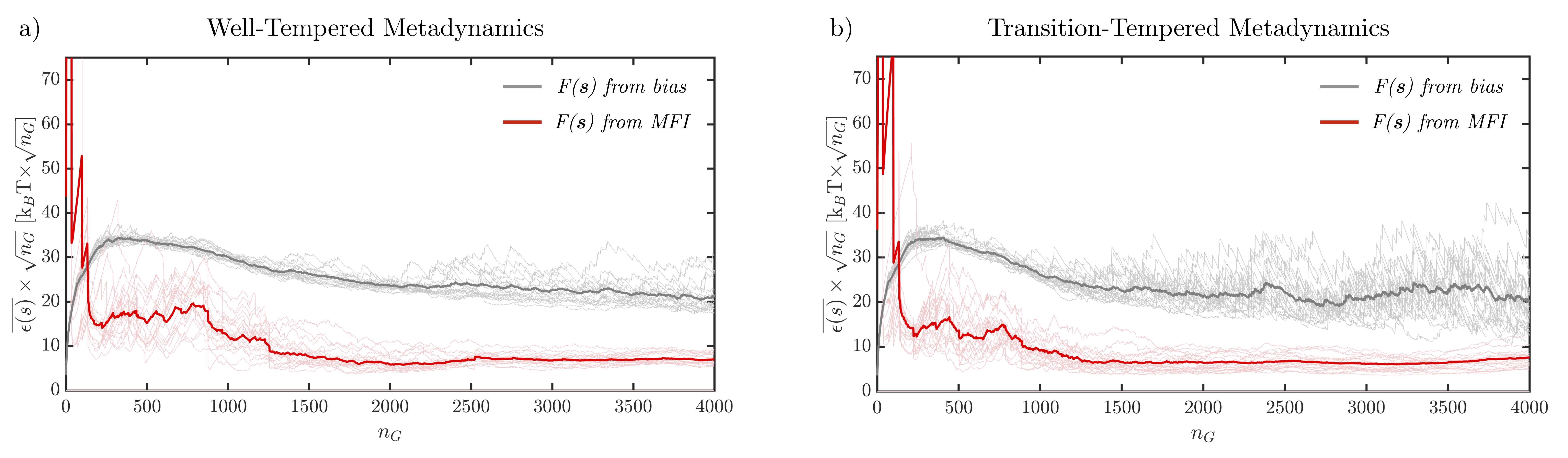}
\caption{Convergence analysis of MFI (red) against standard estimators of the free energy (grey) for a double well model potential. Analysis performed on a well-tempered (a) and transition-tempered (b) metadynamics simulations.}
\label{fig:1Dpotential-additional}
\end{figure*}

Fig.\ref{fig:1Dpotential-additional} reports the accuracy and convergence rates of the 1D model potential obtained by MFI for Langevin dynamics performed with WTmetaD and TTmetaD algorithms. 

\section*{Ibuprofen Conformational Isomerism in the crystal bulk}
\subsection*{Simulations Setup}

The conformational landscape of ibuprofen in the crystal bulk was recovered from MD simulations with the aid of WTmetaD and the resulting FES was used as a reference in the validation of the MFI method. 

MD simulations were performed using Gromacs 5.1.4  \cite{VanDerSpoel2005}, with the Generalised Amber Force Field (GAFF) \cite{Wang2004}. WTmetaD and collective variable post-processing have been carried out using Plumed 2.4 \cite{tribello2014plumed}.
The system discussed in this study refers to a crystal supercell of 3.1$\times$4.3$\times$7.4 nm containing 320 ibuprofen molecules. 
Prior MD the crystal supercell undergoes an energy minimisation with a conjugate gradient algorithm with a tolerance on the maximum residual force of 100 kJ mol$^{-1}$ nm$^{-1}$. 
In MD simulations a cut-off of 1.0 nm for non-bonded interactions is chosen, three-dimensional periodic boundary conditions (\textit{pbc}) are applied and long range intermolecular interactions are accounted for using the particle-mesh Ewald (PME) approach \cite{Darden1993}. Covalent bonds are constrained using the LINCS algorithm \cite{Hess1997}, which enables the use of a 2 fs time step for the numerical propagation of the dynamics. 
All simulations were carried in the isothermal-isobaric (NPT) ensemble at pressure of 1 bar and temperature of 300 K, using Bussi-Donadio-Parrinello thermostat \cite{Bussi2007} and a fully anisotropic Berendsen barostat \cite{Berendsen1984}. 
Representative input files for these simulations are available on PLUMED-NEST (www.plumed-nest.org) under the entry plumID:19.071\cite{NEST}.

\begin{figure}[!h]
\includegraphics[width=0.45\linewidth]{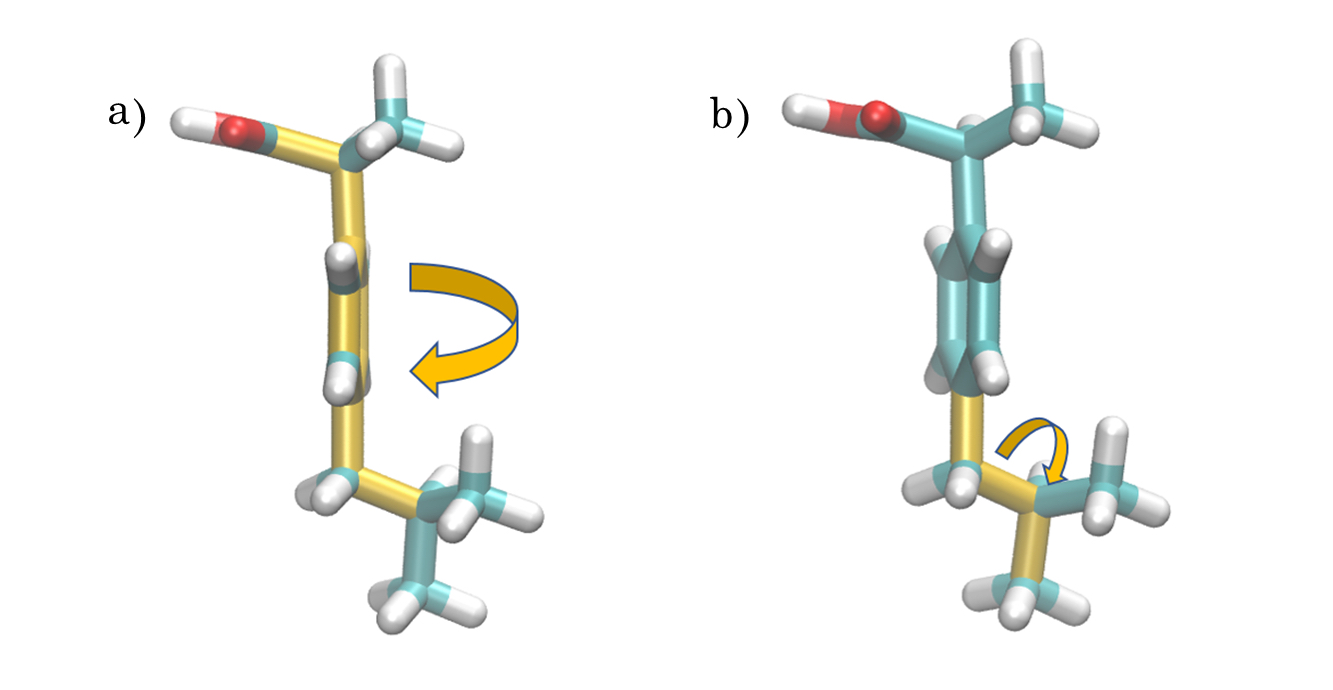}
\caption{a) Global torsional angle monitoring the position of the \textit{para}- substituents of the phenyl ring with respect to the central axis os the molecule (\textbf{s1}) b) Local torsional angle describing the inversion of the two methyl substituents in the isobutanyl substituent (\textbf{s2}). }
\label{fig:Ibuprofen_CVs}
\end{figure}

In the WTmetaD simulations, two collective variables, referred to as \textit{local} and \textit{global} torsional angles, describing the conformational flexibility of a single ibuprofen molecule were used as shown in Fig. \ref{fig:Ibuprofen_CVs}. 

The conformational ensemble of ibuprofen in the crystal bulk is found to be dominated by one conformer referred to as c1. Rotation of the local torsional angle generates conformational isomers c2 and c3, while rotation along the global torsional angle of each of c1, c2 and c3 results in conformers c4, c5 and c6 respectively. We refer to conformers c1, c2 and c3 as the \textit{trans} conformers as the \textit{para}-substituents of the phenyl ring are on opposite sides with respect to the central axis of the molecule, while conformers c4, c5 and c6 form the group of the \textit{cis} ibuprofen conformers. In the crystal bulk conformer c6 of ibuprofen is a high energy state for which the free energy region is not displayed, for more information please refer to Ref. 11 of the main text.

The parameters for the WTmetaD biasing protocol used to generate the reference FES are reported in Tab.\ref{tab:metad_table}. The table also includes the parameters used to generate the independent simulations of conformer transitions. 
\begin{table}
\caption{Parameters in WTmetaD simulations, used to obtain ibuprofen free energy surface in the crystal bulk and independent conformational transitions.}
	\begin{tabular}{ c | cccc }
		& \textbf{Width} & \textbf{Height} & \textbf{Bias Factor} & \textbf{Pace} \\
		& [ rad ] & [$k_BT$] & [K] & [steps] \\
		\hline
		FES &  0.1 & 0.96 & 10 & 500  \\		
		Ind &0.1 & 0.24 & 5 & 1500 \\ 
	\end{tabular}
    \label{tab:metad_table}
\end{table}

\subsection*{Conformational landscape from WTmetaD}
\begin{figure*}[ht]
\includegraphics[width=0.75\linewidth]{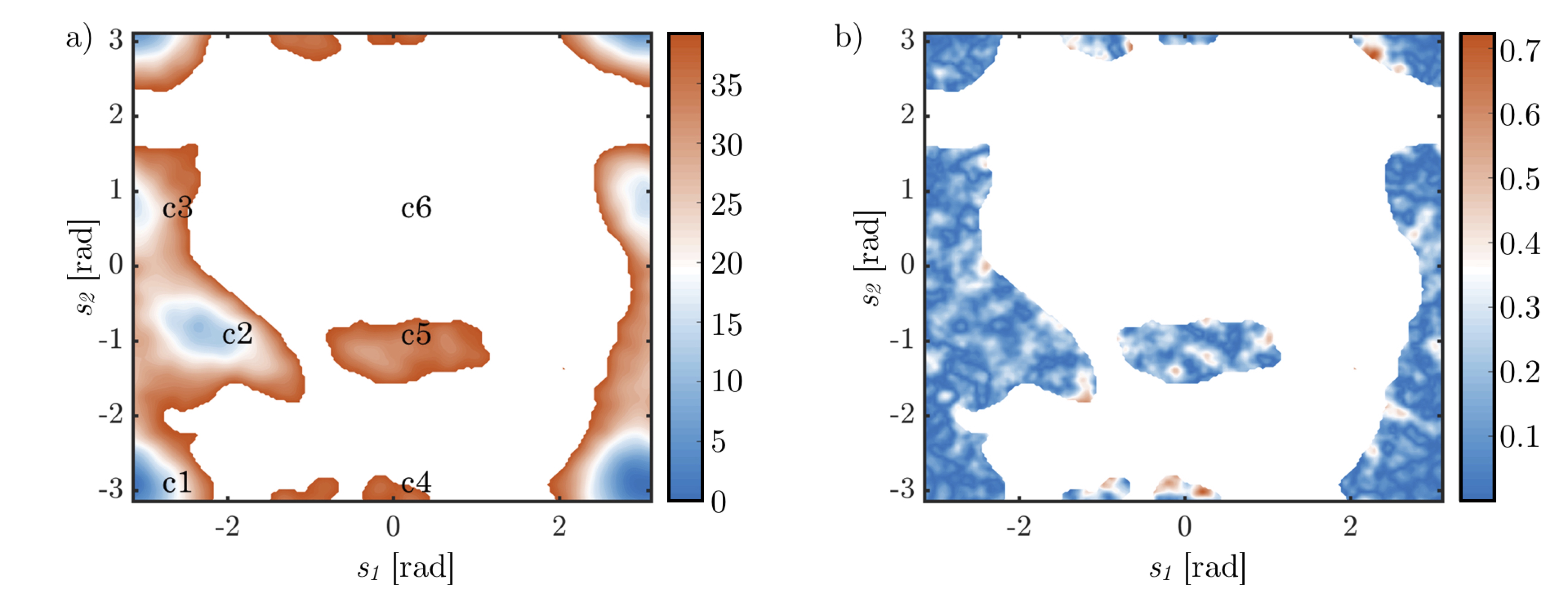}
\caption{a) Free energy profile of the conformational landscape of ibuprofen in the crystal bulk recovered from WTmetaD. b) Position-dependent error of the free energy reported in a)}
\label{fig:Ibuprofen_wtmetad}
\end{figure*}

In Fig. \ref{fig:Ibuprofen_wtmetad}a) we report the free energy surface of the conformational landscape of ibuprofen obtained with WTmetaD along with its associated position-dependent error Fig. \ref{fig:Ibuprofen_wtmetad}b). The error was calculated by the generation of histograms of the trajectory in CV space reweighted with respect to the total biasing potential applied to the system with the WTmetaD protocol. The histograms were generated at an interval of 60 ns and the average standard error was calculated to be 0.28 kJ/mol.

\begin{figure*}[ht]
\includegraphics[width=0.9\linewidth]{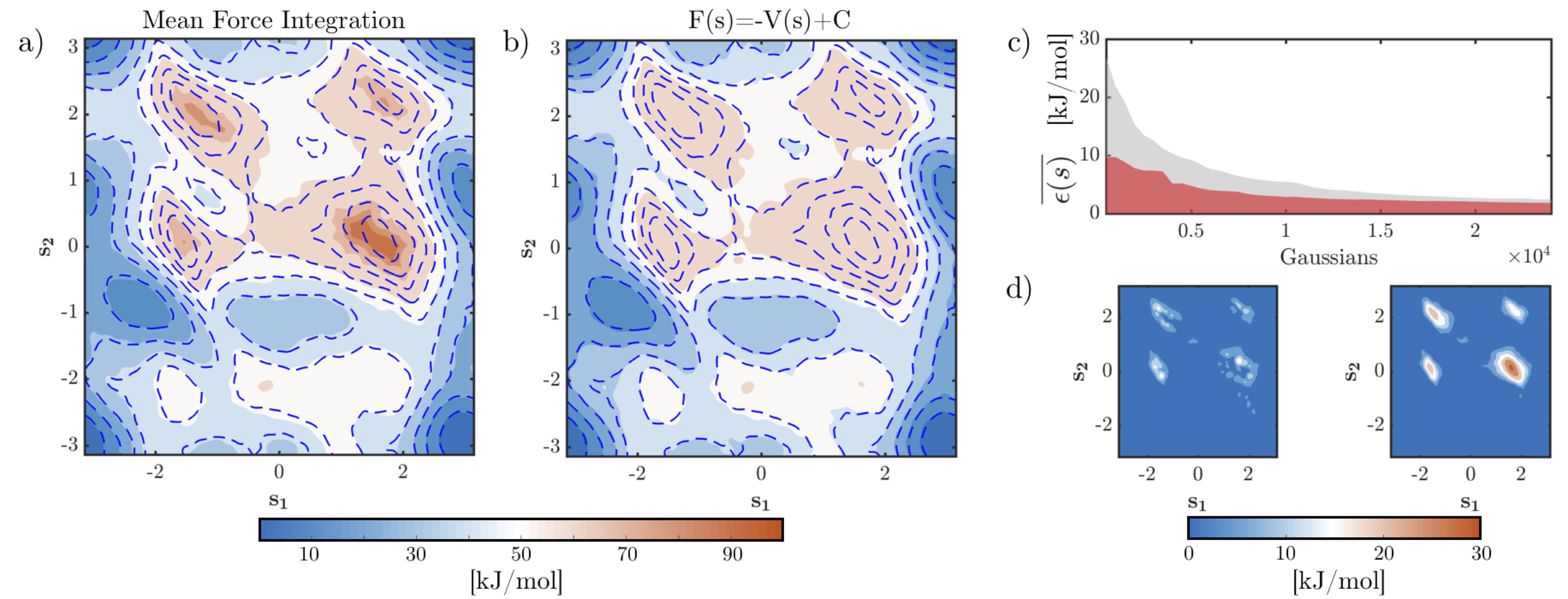}
\caption{a) Free energy surface obtained from MFI b)Free energy surface obtained from the integration of the bias potential. Overlaid isocontours are the $F_{ref}(\mathbf{s})$  obtained from WTmetaD. c) Mean absolute error d) Absolute error in CV space}
\label{fig:Ibuprofen_MFI}
\end{figure*}

We apply MFI to reproduce the conformational free energy landscape of ibuprofen in its crystal bulk. In Fig. \ref{fig:Ibuprofen_MFI} we report the result obtained from MFI (panel a) and from the integration of the bias potential (panel b) after the deposition of 2$\times$10$^4$ Gaussians. To assess accuracy, we define as a reference $F_{ref}(\mathbf{s})$ the FES obtained after 120 ns from WTmetaD. $F_{ref}(\mathbf{s})$ is overlaid to both MFI and total bias estimates as dashed blue isocontours. We report the mean absolute error as a function of simulation time in Fig. \ref{fig:Ibuprofen_MFI} c), showing that MFI displays a higher accuracy at short times. Furthermore, in panel d), we report maps of the absolute error in CV space showing that, given the same sampling time, MFI is more accurate than total bias in reproducing high free energy regions. 

\section{References}
\bibliographystyle{unsrt}
\bibliography{refs.bib}